\documentclass[showpacs,pra]{revtex4}
\usepackage{hyperref}
\usepackage{graphicx}

\def\qpm{_{\mathrm{QPM}}}

\begin{document}

\title{Generation of bright squeezed light at 1.06 $\mu m$ using
  cascaded non-linearities in a triply resonant c.w. PPLN OPO}
\author{K.S. Zhang}
\author{T. Coudreau \footnote{To whom correspondance should be
    adressed : coudreau@spectro.jussieu.fr}}
\author{M. Martinelli}
\author{A. Ma\^\i tre}
\author{C. Fabre} 
\affiliation{Laboratoire Kastler Brossel \\
Universit\'{e} Pierre et Marie Curie, case 74 \\
75252 Paris cedex 05 France}
\date{\today}
\begin{abstract}
We have used an ultra-low threshold continuous-wave Optical Parametric
Oscillator (OPO) to reduce the quantum
fluctuations of the reflected pump beam below the shot noise limit. The
OPO consisted of a triply resonant cavity containing a
Periodically-Poled Lithium Niobate crystal pumped by a Nd:YAG
laser and giving signal and idler wavelengths close to
2.12 $\mu m$ and a threshold as low as 300 $\mu W$. We detected the
quantum fluctuations of the pump beam reflected by the OPO using a
slightly modified homodyne detection technique. The measured noise
reduction was 30 \% (inferred noise reduction at the output of the OPO
38 \%).
\end{abstract}


\maketitle
\section{Introduction}

Continuous-wave OPOs have been used for over a decade to study and
modify the quantum noise of light. Successful experiments involved
creation of squeezed light below the oscillation threshold 
\cite{polzik,bruckmeier}, sub shot noise intensity correlations
between signal and idler beams \cite{heidmann,mertz} and pump
squeezing \cite{kasai}. These experiments have been performed
using OPOs pumped with visible light, generally the second  harmonic
of Nd:YAG lasers, and most of them generated non-classical light at a
wavelength different from the pump wavelength.
Pump squeezing in an OPO above
threshold \cite{kasai} provides a direct and efficient 
way to generate bright squeezed beams without frequency changes. The
OPO acts as a ``quantum noise eater'', like a Kerr medium, by making
use of the so-called ``cascaded nonlinearities'' \cite{cascade}. Such
cascaded non linearities, resulting from the co-existence of sum frequency
generation and parametric down  conversion occuring in an OPO, have
been shown to amount to a nonlinear phase-shift on the pump beam
which emerges from the cavity after its nonlinear interaction with the
signal and idler beams in the parametric crystal. This effect has been
known for a long time to be able to produce significant quantum noise
reduction \cite{reynaud}. In order to have access to a new frequency range
in pump squezing, we have used a Quasi Phase-Matched (QPM) material,
Periodically-Poled Lithium Niobate (PPLN), which makes it possible to build
low-threshold c.w. OPOs operating with a Nd:YAG laser as a pump and
signal and idler beams in the mid-IR
\cite{martinelli,byer,schiller}. 
The experiment that we present here is the first,
to the best of our knowledge, to use QPM material for a quantum noise
reduction experiment in the c.w. regime.

After a brief description of the
experimental set-up, we will study in section \ref{meanfields} the
behavior of the mean fields produced by this device and compare it to
theoretical predictions. In the last part, we will
present the theoretical predictions as well as the experimental
results for the quantum noise, showing 38\% noise reduction below the
standard quantum limit.

\section{Experimental set-up}

\begin{figure}
\centerline{\includegraphics[clip=,width=8cm]{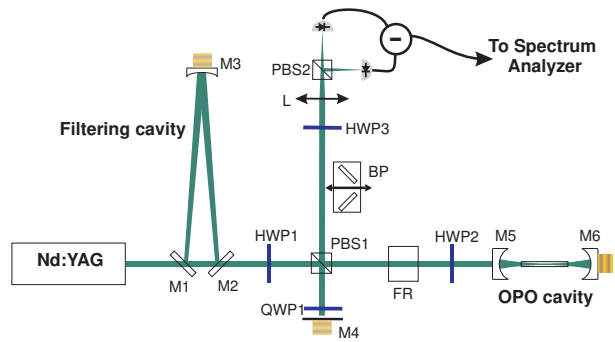}}
\caption{Experimental set-up}
\label{manip}
\end{figure}

\begin{figure}
\centerline{\includegraphics[clip=,width=8cm]{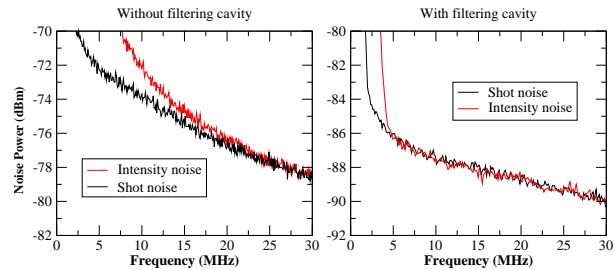}}
\caption{Intensity noise before and after the filtering cavity showing
  that the output beam is shot noise limited above 5 $MHz$}
\label{fc}
\end{figure}

The set-up is shown on Fig. \ref{manip}.
We have used a diode-pumped monolithic Nd:YAG laser (Lightwave
126-1064-700) as a pump source. This source delivers a c.w. beam in a
TEM$_{00}$ mode which exhibits  very large excess noise below 20 $MHz$
due to the relaxation oscillation peak noise of the pump semiconductor
laser (left, Fig. \ref{fc}). In order to obtain noise reduction below
the standard quantum limit within the OPO cavity bandwidth, we need
first to obtain a pump beam at the shot-noise level. For this purpose,
we send the beam through a filtering cavity. It is a ring three-mirror
cavity \cite{filtcav} to prevent back reflections from perturbing the
laser (Fig. \ref{manip}). The input and output mirrors (M1, M2) are
identical plane mirrors with a reflexion coefficient on the order of
99.5 \% for the $s$-polarisation. The end mirror (M3) is a
concave mirror with a radius of curvature of 750 $mm$ and a reflexion
coefficient close to 99.9 \%. The measured finesse for the cavity is
close to 700. For a 70 $cm$ round trip length the cavity
bandwidth is below 1 $MHz$ when the cavity is locked on resonance.
The field fluctuations of the transmitted beam are
shot-noise limited above  5 $MHz$ (right, Fig. \ref{fc}).
 Due to the non negligible losses on the mirrors, we obtain 30 \%
power transmission as opposed to 80 \% theoretical prediction and 75
\% in similar experiments where very-high quality mirrors were used
\cite{filtcav}. When the cavity is locked by a servo loop using a
frequency modulation of the Nd:YAG laser at 40 kHz, its output is
stable over several tens of minutes.  
The mode matching  of the pump beam to the OPO cavity was 92\% 
without the
filtering cavity and over 97 \% with the filtering cavity which plays
a role of spatial filter.

The half wave plate HWP1 and the polarizing beam splitter PBS1
control the pump intensity sent into the OPO cavity.
The pump beam passes through
the Faraday Rotator (FR), and the input polarization is
controled by the half wave plate HWP2. The reflected pump beam, after the
interaction with the OPO cavity, returns by the same optics and is reflected by
PBS1 after the double passage through the Faraday Rotator.
For the local oscillator, we used part of the input beam reflected by
PBS1. The reflected beam passes by an optical circulator made by a quater wave
plate (QWP1) and a mirror (M4) mounted in a piezo-actuator for the phase
variation. The crossed polarized local oscillator and reflected pump
beams are then mixed at
the homodyne detection system by the polarizing beam splitter PBS2 and the half
waveplate HWP3, and detected by the high efficiency InGaAs photodetectors
(Epitaxx ETX 300, quantum efficiency 94 \%).
A pair of Brewster Plates (BP) at the Brewster angle for
the local oscillator polarization can
be introduced in the system, producing a
reduction in the reflected pump beam with no noticeable change
in the local oscillator.

The triply resonant OPO has been described
extensively in a previous paper \cite{martinelli} and we will only
recall here its main properties. 
The cavity is a symmetric cavity with a length of  
approximately 65 $mm$. The pump coupling mirror
(M5) has a reflection coefficient of 87 \%, so that the transmission is much
larger than the other losses inside the cavity. 
The end mirror (M6) is highly reflecting
for the pump beam to reduce the losses (reflection of 99.8 \%).
The mirrors have large reflection coefficients
in the 2 $\mu m$ range to ensure small losses for the pump beam due to
conversion to signal and idler (99.8 \% for M5, 99 \% for M6). They have a 
radius of curvature of 30 $mm$ to ensure optimal waist sizes in the
crystal (36 $\mu m$ at 1.06 $\mu m$, 51 $\mu m$ at 2.12 $\mu m$).

The crystal is a PPLN crystal from Crystal Technologies, with a width
of 12 $mm$, a  length of 19 $mm$ and a thickness of 0.5 $mm$. Both
faces of the crystal are antireflection coated for pump, signal and idler
frequencies (residual reflection of 0.6 \% for the pump and 0.4 \%
for signal and idler). The crystal absorption in the infrared region is
small (0.3 \% at 1.06 $\mu m$ according to the manufacturer).
The crystal is formed of 8 different paths with spatial
periodicity varying between 30 and 31.2 $\mu m$. All the experimental
results shown in this paper were made using a 31.1 $\mu m$ spatial
period for which the exact quasi-phase matching condition for
degenerate signal and idler wavelength at 2.12 $\mu m$ is obtained
with a temperature $T \qpm \approx 162 ^\circ C$. The crystal is placed
inside a temperature stabilized oven. In presence of the crystal, the
measured finesse for the pump is around 40, in good agreement with the
theoretical expressions while the calculated finesse for signal and idler is
around 200. The typical oscillation threshold is around 500 $\mu W$
with thresholds as low as 300 $\mu W$ for a limited time. For continuous
operation, the OPO cavity is locked in the resonance of the 2 $\mu m$ output.

\section{Mean fields study} \label{meanfields}

The well-known equations for the intracavity mean fields are \cite{sozopol}
\begin{eqnarray}
E_0 &=& t_0 E_{in} + r_0 (E_0 - \chi^\ast E_1 E_2) e^{i \varphi_0}\\
E_1 &=& r (E_1 + \chi E_0 E_2^\ast) e^{i \varphi_1} \label{eqsignal}\\
E_2 &=& r (E_2 + \chi E_0 E_1^\ast) e^{i \varphi_2} \label{eqidler}
\end{eqnarray}
where $E_0$, $E_1$ and $E_2$ are the intracavity pump, signal and
idler fields respectively, $t_0$ and $r_0$ are the amplitude
transmission and reflection coefficients at the pump frequency, $\chi$
is the nonlinear coupling coefficient, $r$ is the common value for the
amplitude reflection coefficient at the signal and idler
frequencies. $\varphi_0$, $\varphi_1$ and 
$\varphi_2$ are the round-trip phases for pump, signal and idler
fields respectively. Above threshold,
equations (\ref{eqsignal}) and (\ref{eqidler}) lead to the condition
\cite{debuisschert}
\begin{equation}
\varphi_1 = \varphi_2 + 2 p \pi \label{eqphi}
\end{equation}
where $p$ is an integer. Each value of $p$ is associated with a given
mode, \emph{i.e.} a well-defined couple of signal and idler
frequencies, $\omega_1$ and $\omega_2$. More precisely, using the
Sellmeier equation for lithium niobate \cite{jundt} and
eq. (\ref{eqphi}), one can find for any given cavity length
$L_{cav}$, crystal temperature $T$ and mode index $p$ the values of
$\omega_1$ and $\omega_2$. $T$ and $p$ being fixed, a minimum of the
oscillation threshold is
reached at values of the cavity length where both $\varphi_1$ and
$\varphi_2$ are multiple of $2 \pi$. Let us call $L_p$ the
signal-idler double resonance point within a given free spectral
range.

In order to calculate the oscillation threshold and output fields, we
have taken into account the periodical poling. The usual phase
mismatch, $\Delta k = k_0 - k_1 -  k_2$, is replaced by $\Delta \kappa
= k_0 - k_1 - k_2 - 2 \pi / \Lambda$ where $\Lambda$ denotes the
spatial period of the crystal. The nonlinear coupling coefficient which
can be written in a usual bulk crystal 
\begin{equation}
\chi = \chi^{(2)}_{bulk} = \chi^{(2)} (\omega_0,
\frac{\omega_0 }{2}, \frac{\omega_0 }{2} ) \mathrm{sinc} \left(
\frac{\Delta k l}{2} \right) \exp \left(- i \frac{\Delta k l}{2} \right)
\end{equation}
now becomes
\begin{equation}
\chi = \chi^{(2)}_{QPM} = \chi^{(2)} (\omega_0,
\frac{\omega_0 }{2}, \frac{\omega_0 }{2} ) \frac{2}{\pi} \mathrm{sinc}
\left( \frac{\Delta \kappa l}{2} \right) \exp \left(- i \frac{\Delta
\kappa l}{2} \right)
\end{equation}
for an integer number of periods within the crystal of lenght $l$.
 If the crystal
does not consist of an integer number of periods, the coupling
coefficient has an additionnal term like $\chi^{(2)}_{bulk}$,
depending on the added fraction of period. Since the relative
contribution of this term will have the order of  $\Lambda$/L, it can
be neglected in the calculation of $\chi^{(2)}_{QPM}$.
 However, its contribution to the crystal length 
will have a much more important effect in the relative
phase-shift between the interacting beams. The effect of such
an additional phase-shift is known to increase the threshold
in the case of a linear OPO cavity \cite{debuisschert}.
 This increase will depend on this phase term, reaching 1.9 times the mininum
 threshold that would be obtained with a perfect phase matching, perfect
periodicity of the crystal and no added relative phase.
 In order to optimize
the threshold, it is possible to use crystals with non parallel faces
which allow a precise choice of the crystal length \cite{kamel,fejer}.

\begin{figure}
\centerline{\includegraphics[clip=,width=12cm,angle=-90]{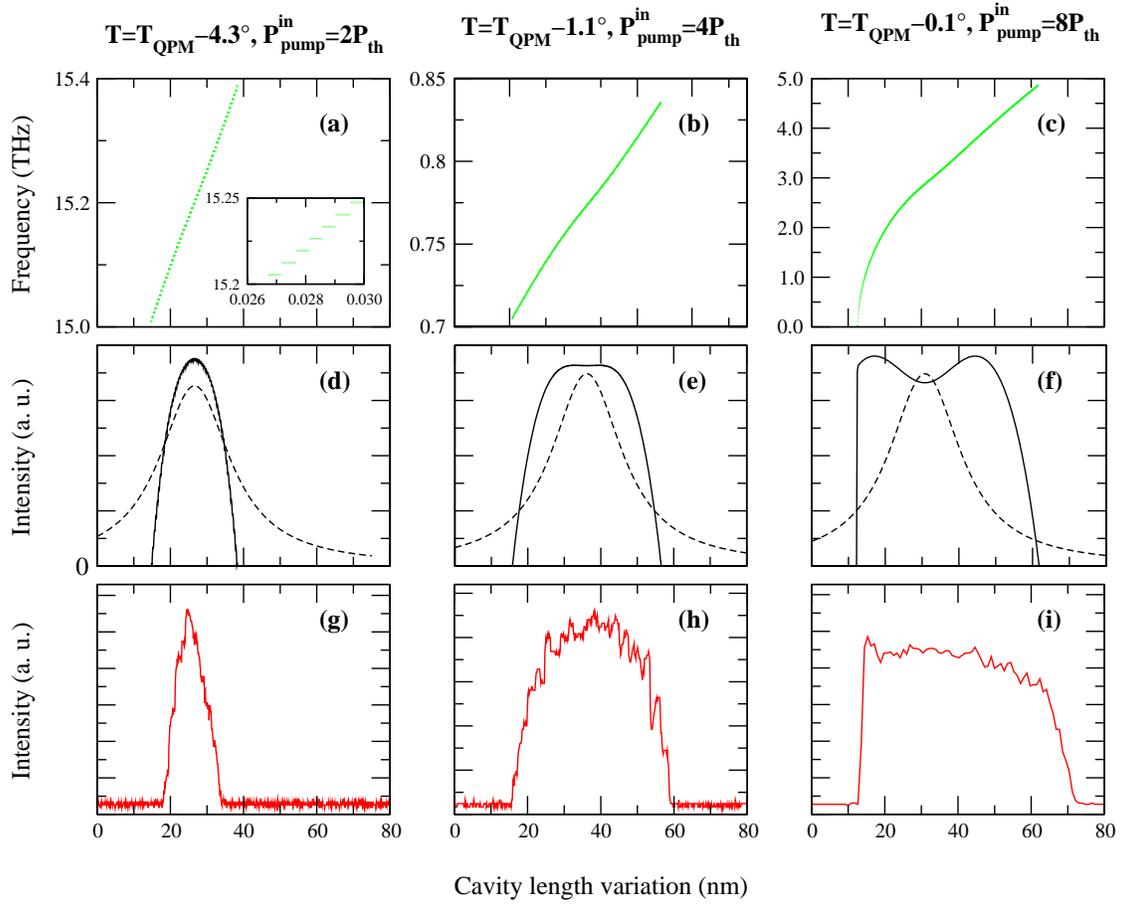}}
\caption{OPO mean values study as a function of the variation of cavity
  length. The frequency difference between signal and idler in
  $THz$ is shown on traces (a), (b) and (c).
  Calculated signal mean intensities (traces (d), (e), (f),
  continuous line) and experimental curves (traces (g), (h) (i))
  are shown with the same scales; the undepleted pump mean intensity is
  shown for comparison (dashed lines). The horizontal axis is in $\mu m$.  
  The experimental conditions (temperature and pump power) are:
  $T=T \qpm - 4.3 ^\circ$ and  $P_{pump}^{in} = 2 P_{th}$ (traces (a), (d)
  and (g)), $T=T \qpm - 1.1 ^\circ$ and $P_{pump}^{in} = 4 P_{th}$ 
  (traces (b), (e) and (h))
  and $T=T \qpm - 0.1 ^\circ$ and $P_{pump}^{in} = 8 P_{th}$ (traces
  (c), (f) and (g)).
  $P_{th}$ is  the minimum threshold obtained at exact resonance for pump,
   signal and idler at a given temperature. In curve (a) the zoom area shows
   the mode-hop during the cavity scan.} 
  \label{chpm}
\end{figure}

Curves (d), (e) and (f) on Fig. \ref{chpm} show the mean field
intensities calculated from these equations as a function of cavity
length for three different temperatures and pump intensities.
For a given cavity length, there are many different modes $p$ 
where the OPO can oscillate. For each one of these modes there will be 
a different threshold value, and the OPO will operate in the mode with
the lowest threshold \cite{schwob}. The calculated
frequency difference between signal and idler is shown in $THz$ on
traces (a), (b) and (c) showing that the system can be swept over a large
frequency range. During the cavity scanning, the OPO passes through  a
large number of individual modes, remaining on a given value of the
frequency difference only over a limited range of cavity length
(inset, curve (a)). When the temperature approaches $T \qpm$, the
signal and idler frequencies become very close, and the difference
between signal and idler indices of refraction becomes very small. In
these conditions, the resonance lengths $L_p$ of different oscillation
modes $p$ become very close to each other and the possible oscillating
modes overlap. As the OPO is never multimode in steady state operation
\cite{schwob,cras}, it jumps from one mode to the next ($p \rightarrow
p \pm 2$) when the cavity 
length is varied. As a result, the OPO always oscillate very close to
the double resonance configuration. The mode jumps can be seen as
discontinuities on the mean field. These discontinuies are easily visible
on the experimental curves and using an auxiliary Fabry-Perot, we have
observed monomode operation with mode hops at the
discontinuities. These discontinuities are not seen on the calculated
curves which do not take into account the dynamics of the
system while experimentally, since the threshold varies slowly for the
different modes, the OPO remains on its initial mode even if it is
unstable, leading to mode jumps by more than two units.

Intracavity signal and idler intensities
exhibit a bistable behaviour as a function of cavity length when
specific conditions on the pump and signal detunings are met
\cite{lugiato}. In our case, as the OPO works almost always with a small
detuning of the signal and idler fields, it is difficult to
observe any bistable behavior. Nevertheless, due to the parabolic shape of
the phase-matching curve in a type-I medium \cite{eckardt}, for a
temperature close to $T\qpm$, oscillation is limited by the frequency
degeneracy of signal and idler ($p=0$). Thus the
detuning range obtained for the mode $p=0$ is much larger than that
obtained for other modes. Such a behaviour may be seen on the left
hand side of curve (f). In this region, for an adequate cavity
detuning the bistable behavior is expected. The value of the pump
detuning can be selected by the temperature control of the phase
matching condition at degeneracy, displacing the position of the sharp
side of the signal and idler curve relative to the  pump resonance
position. The sharp edge position is critically dependent on the
temperature, and a $ 0.1 ^\circ C$ variation is enough to displace the
degenerate point out of the pump resonance.

\section{Quantum fluctuations of the pump field}

\begin{figure}
\centerline{\includegraphics[clip=,width=8cm,angle=-90]{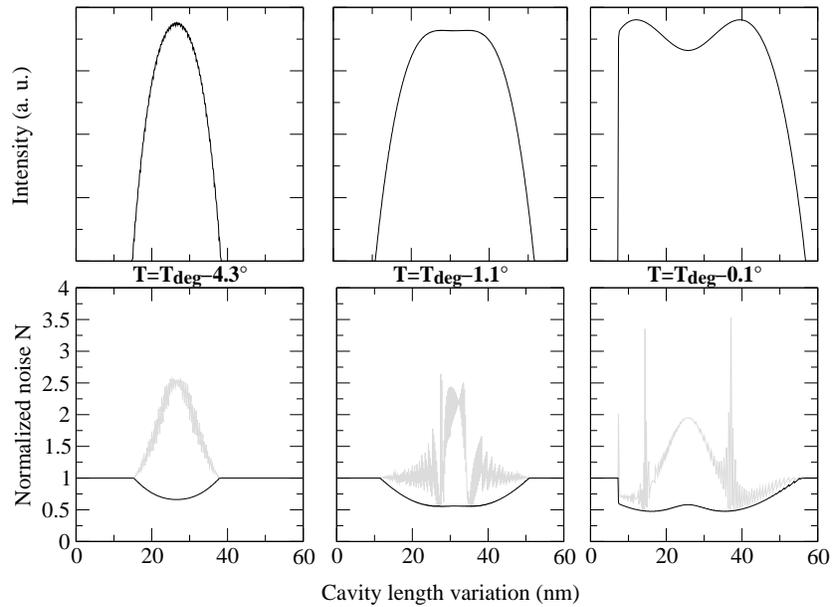}}
\caption{Calculated mean intensities (upper traces) and noise (lower
  traces, dark line : optimum noise reduction, grey line : intensity
  noise ) as a function of cavity length in the same conditions as Fig. 3.}
  \label{sq}
\end{figure}

The quantum fluctuations of the pump field have been calculated in
Ref. \cite{lugiatosemclas}. This paper shows that 
best quadrature squeezing on the pump field increases with pump power
and is already significant a few times (by a factor 3 or 4) above
threshold and for any value of the detunings. It increases when
approaching the bistability threshold, i.e. for rather important
signal-idler and pump detunings. Squeezing occurs on the phase
component at exact double resonance \cite{kasai} and,
for non-zero detunings, on a quadrature component which rotates quickly when
this detunning changes. Using the detailed
calculations of \cite{lugiatosemclas}, it is possible to 
evaluate the optimum noise reduction as a function of cavity length at
a given temperature. The curves are shown on Fig. \ref{sq}. It can be
seen on these curves that the optimum squeezing is significant
on a broad range of detunings and increases with the
pump power.
On the other hand, the intensity noise (grey line) exhibits rapid oscillations
for small changes of the cavity length (of the order of tenths of nm).
Even with a stable electronical locking of the OPO cavity, fast small
variations around the locking position cannot be avoided, thus reducing the
measurable intensity squeezing.
Actually the most efficient locking is obtained
exactly at the triply resonant condition when only quadrature squeezing can be
observed.

The quantum noise of quadrature components of the reflected pump beam are
measured using a slightly modified homodyne detection procedure. In
order to prevent saturation of the detectors (which occurs around 3
$mW$), the local oscillator cannot be much more powerful than the 
reflected pump beam. The difference of the
photocurrents fluctuations obtained by the two balanced photodetectors
can be written as
\begin{equation}
\Delta^2( i_1 -  i_2) \propto I_{LO} \Delta^2 E_{\theta}
\end{equation}
only when the local oscillator is much stronger than the beam to measure.
$I_{LO}$ denotes the mean local oscillator power and $\Delta^2
E_{\theta}$  the variance of the reflected pump in the quadrature
$\theta$ determined by the phase difference between the local oscillator
and the reflected pump mean field. The measurement, with a spectrum
analyzer, of the spectral density of the difference of the
photocurrents gives directly the noise spectrum of the quadrature
component of angle $\theta$. 
In the general case of finite ratio between the mean intensities of
the beam to study and of the local oscillator beam, an additionnal
term appears and the difference of the photocurrents is now 
\begin{equation}
\Delta^2( i_1 - i_2 )\propto I_{LO} \Delta^2 E_{\theta}+I \Delta^2 E_{LO,\, p}
\label{ditotal}
\end{equation}
where we neglect correlations between the two beams (this is justified
since these beams are obtained by splitting a coherent state on a beam
splitter). $I$ is the mean reflected pump intensity and $\Delta^2
E_{LO,\, p}$ the variance of the pump beam in a given quadrature. This
extra term can be measured independently. Taking into account the fact
that the local oscillator is in a coherent state, its high frequency
noise is equal to that of the vacuum noise and can be measured by
blocking the local oscillator arm. Using this measurement, we can
renormalize the total photocurrent difference (\ref{ditotal}) and
obtain the reflected pump noise. This correction amounts to
approximately 30 \% of the shot noise.

\begin{figure}
\centerline{\includegraphics[clip=,width=8cm,angle=-90]{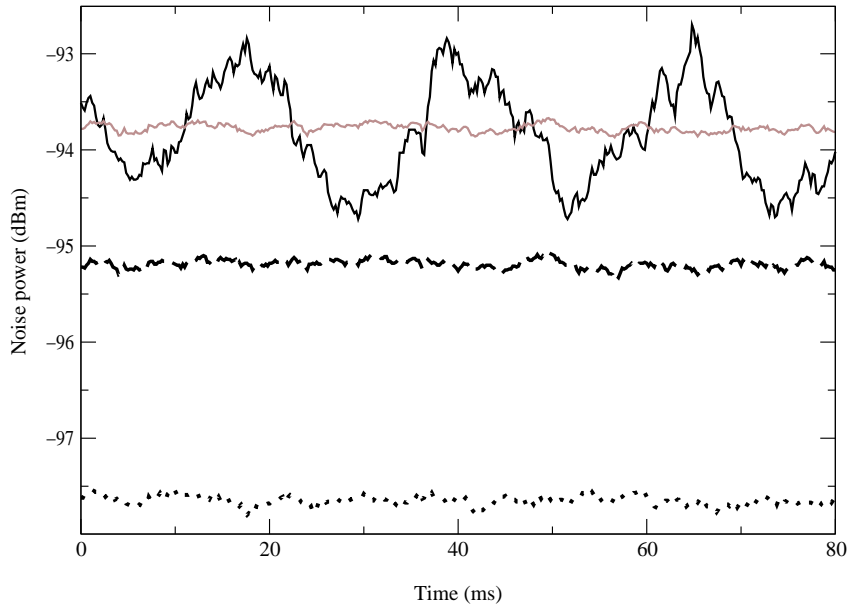}}
\caption{Noise power measured at the homodyne detection.
  The thick black line is obtained with both local oscillator and
  reflected pump during the scan of the local oscillator phase.
  The dashed line is the local
  oscillator shot noise level, obtained for a blocked pump.
  The dotted line is the shot noise level for the reflected pump
  (blocked local oscillator).
  The grey line is the calculated shot noise level obtained from the sum
  of the local oscillator and the equivalent reflected pump shot noise power
  and corresponds to the expected shot noise with the two beams present (since
  the beams have uncorrelated fluctuations).
  The electronic background noise level is -102.6
  dBm. The experimental conditions are the 
  following~: $P_{pump}^{in} = 1.2 mW$, $P_{pump}^{det} = 0.45 mW$,
  $P_{LO} =1.2 mW$. The experimental settings of the spectrum analyser are:
  noise analysis frequency $= 6 MHz$, resolution bandwidth $= 100 kHz$
  and video bandwidth $= 10 kHz$}  
\label{sqraw}
\end{figure}

\begin{figure}
\centerline{\includegraphics[clip=,width=8cm,angle=-90]{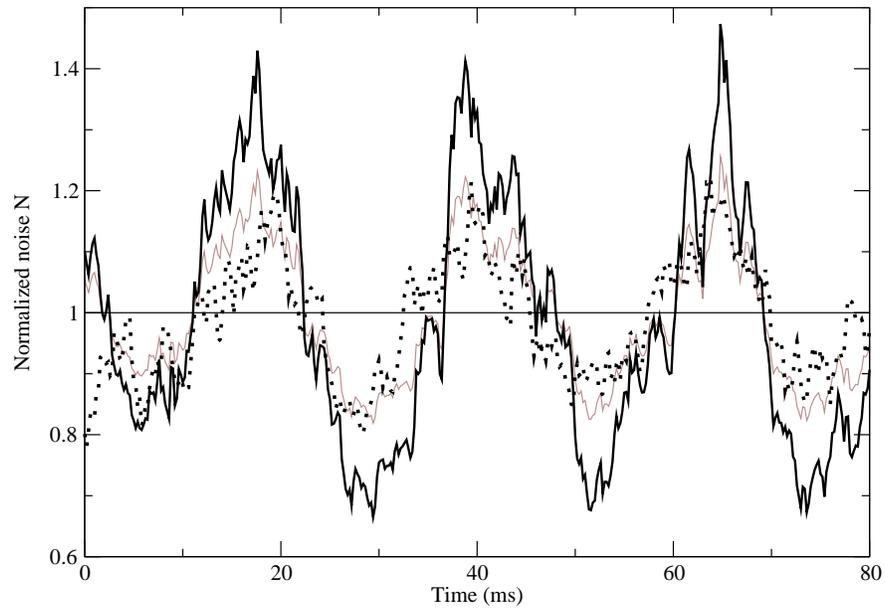}}
\caption{Normalized noise as a function of time when the local
  oscillator phase is scanned. The thick black line is the
  unattenuated beam noise showing 30\% noise reduction below the
  standard quantum limit. The
  dashed line shows the measured noise in the presence of the pair of
  Brewster plates. The grey line is the calculated noise taking into
  account the losses introduced by the pair of Brewster plates.
  The thin black line corresponds to the shot noise limit.}  
\label{sqmes}
\end{figure}

The noise measurement is made with a pair of balanced InGaAs
photodetectors. The mode
matching  between the reflected pump and the local oscillator is on
the order of 97 \%. The pump intensity incident on the OPO is
$P_{pump}^{in} = 1.2 mW$, for a threshold power of 300 $\mu$W. The
reflected pump intensity at the position of the photodetectors was
$P_{pump}^{det} = 0.45 mW$. The temperature 
is $T = T \qpm -1^\circ C $ and the local oscillator power is $P_{LO}
=1.2 mW$.
Figure \ref{sqraw} shows the noise power $N _{1}$ obtained as the local
oscillator phase is scanned (thick black line), as well as the shot noise
level measured for the local oscillator and the reflected pump
 beams (respectively $N _{2}$ and $N _{3}$). 
The sum of the shot noise power of both beams is also presented for
comparison. The electronic noise level is
small (-102.6 $dBm$) and is taken into account in our calculations.
The variance of the quadrature component fluctuations, normalized to the
coherent state fluctuations
($N = \Delta^2 E_{\theta}/\Delta^2 E_{LO,\, p}$) can be obtained from the
noise measurements shown at Fig. \ref{sqraw} using the
correction presented in eq. (\ref{ditotal}).
Figure \ref{sqmes} shows this normalized noise $N = (N _{1}-N _{3})/N _{2}$.
The measured noise reduction is 30 \%
corresponding to an inferred noise reduction at the output of the OPO
of 38 \%, when losses are taken into account.

To check that we have indeed noise reduction below the
standard level, we have introduced losses on the squeezed beam. In
order for these losses not to change the operating conditions (namely
the incident pump power, $P_{pump}^{in}$), we have introduced
losses using a pair of glass plates oriented at the Brewster angle with
respect to the propagation axis (Fig. \ref{manip}): in that case,
the losses are 46 \% for the squeezed beam and not measurable for the
local oscillator since the two beams are orthogonally polarized. 
The dashed grey line of Fig. \ref{sqmes} gives the quadrature noise
measured using this device. This line coincides well with the full
grey line which represents the measured noise without attenuation
(thick black line) corrected using the well-known formula \cite{atenuation}
 $N_{loss}=(1-\Gamma)N+\Gamma$, where $N_{loss}$ is the normalized noise 
of the attenuated beam, $N$ is the normalized noise of the input beam,
 and $\Gamma$ is the loss coefficient of the inserted attenuator.
 
The measured noise reduction is in good agreement with the quadrature
noise reduction obtained in our theoretical calculations
(44 \%, Fig. \ref{sq}).
The smaller value for the measured compression may be due
to an underestimated value for the intracavity losses at our calculations.

\section{Conclusion}

We have reported on the first use of quasi-phase matched materials for
cw quantum noise reduction experiments and observed an inferred
noise reduction of 38 \% 
below the standard quantum limit using such a material. The
use of quasi-phase matched materials in quantum optics seems very
promising as it enlarges the wavelength range where quantum effects
can be obtained.

\begin{acknowledgments}
M.M. wishes to thank Coordena\c c$\tilde\mathrm a$o de Aperfei\c
coamento de Pessoal de N\' \i vel Superior (CAPES-BR) for
funding. This research was performed in the framework of the EC ESPRIT
contract  ACQUIRE 20029.
\end{acknowledgments}

\begin{thebibliography}{99}

\bibitem{polzik} E.S. Polzik, J. Carri and H.J. Kimble, Phys. Rev. Lett.,
  \textbf{68}, 3020 (1992).

\bibitem{bruckmeier} K. Schneider, R. Bruckmeier, H. Hansen,
  S. Schiller, and J. Mlynek, Opt. Lett. \textbf{21}, 1396 (1996).

\bibitem{heidmann} A. Heidmann, R. J. Horowicz, S. Reynaud,
  E. Giacobino, and C. Fabre, Phys. Rev. Lett. \textbf{59}, 2555
  (1987).

\bibitem{mertz} J. Mertz, T. Debuisschert, A. Heidmann, C. Fabre and
  E. Giacobino, Opt. Lett \textbf{16}, 1234 (1991).

\bibitem{kasai} K.~Kasai, JiangRui~Gao and C.~Fabre,
  Europhys. Lett. \textbf{40}, 25 (1997).

\bibitem{cascade} A.G. White, J. Mlynek, and S. Schiller, Europhys.
  Lett. \textbf{35}, 425 (1996).

\bibitem{reynaud} S. Reynaud, C. Fabre, E. Giacobino and A. Heidmann,
  Phys. Rev. \textbf{A 40}, 1440 (1989).

\bibitem{martinelli} M. Martinelli,  K.S. Zhang, T. Coudreau,
  A. Ma\^{\i}tre and C. Fabre, submitted to JEOS \textbf{A}.

\bibitem{byer}  D.~Serkland, R.~Eckardt, R.~Byer, Optics Letters 
  \textbf{19}, 1046 (1994).

\bibitem{schiller} U. Str\"ossner, A. Peters, J. Mlynek, S. Schiller,
  J.-P. Meyn and R. Wallenstein, Opt. Lett. \textbf{24}, 1602 (1999).

\bibitem{filtcav} B. Willke, N. Uehara, E.K. Gustafson, R.L. Byer,
  P. King, S. Seel, R.L. Savage, Jr., Opt. Lett. \textbf{23}, 1704
  (1998).

\bibitem{sozopol} C. Fabre, in \textit{Advanced Photonics with
 Second-order Optically Nonlinear Processes}, edited by A. D. Boardman et al.
  (Kluver Academic Publishers, Netherlands, 1999).

\bibitem{debuisschert} T. Debuisschert, A. Sizmann, E.
  Giacobino, C. Fabre, J. Opt. Soc. Am. \textbf{B 10}, 1668 (1993).
 
\bibitem{jundt} D.H. Jundt, Opt. Lett. \textbf{22}, 1553 (1997).

\bibitem{kamel} I. Juwiler, A. Arie, A. Skliar, G. Rosenman,
  Opt. Lett. \textbf{24}, 1236 (1999).

\bibitem{fejer} G. Imeshev, M. Proctor, M. M. Fejer, Opt. Lett. \textbf{23},
165, (1998).

\bibitem{schwob} C. Schwob, P.-F. Cohadon, C. Fabre, M. Marte,
  H. Ritsch, A. Gatti, L. Lugiato, Appl. Phys. \textbf{B 66}, 685
  (1998).

\bibitem{cras} C. Fabre, M. Vaupel, N. Treps, P.-F. Cohadon,
  C. Schwob, A. Ma\^{\i}tre, C. R. Acad. Sci. Paris, t.1, S\'erie IV,
  553 (2000).

\bibitem{lugiato} L.A. Lugiato, C. Oldano, C. Fabre, E. Giacobino,
  R.J. Horowicz, Nuov. Cim. \textbf{10}, 959 (1988).

\bibitem{eckardt} R. C. Eckardt, C. D. Nabors, W. J. Kozlovky, R. L. Byer,
J. Opt. Soc. Am. \textbf{B 8}, 646 (1991).

\bibitem{lugiatosemclas} C. Fabre, E. Giacobino, A. Heidmann, L. 
  Lugiato, S. Reynaud, M. Vadacchino, Wang Kaige, Quantum Opt.,
  \textbf{2}, 159, (1990).

\bibitem{atenuation} G. Grynberg, A. Aspect, C. Fabre, \textit{Introduction
 aux Lasers et \`{a} l'Optique Quantique} (Ed. Ellipses, Paris, 1997).

\end{thebibliography}
\end{document}